\RequirePackage{lineno} 
\documentclass{emulateapj}
\usepackage{times}
\usepackage{graphicx}

\slugcomment{To appear in the Astrophysical Journal}

\shorttitle{Sub-luminous $\gamma$-ray Pulsars}
\shortauthors{Romani et al.}

\begin{document}

\title{Sub-luminous ${\gamma}$-Ray pulsars}

\author{R. W. Romani, M. Kerr\altaffilmark{1} and H. A. Craig}
\affil{Department of Physics, Stanford University, Stanford, CA 94305} 
\author{S. Johnston}
\affil{Australia Telescope National Facility, CSIRO, Epping, NSW 1710, Australia}
\author{I. Cognard}
\affil{Laboratoire de Physique et Chimie de l'Environnement, LPCE UMR 6115 CNRS, 45071 Orl\'eans
Cedex 02, and Station de radioastronomie de Nan\c cay, Observatoire de Paris, CNRS/INSU, 18330 Nan\c cay, France}
\author{D.A. Smith}
\affil{Universit\'e Bordeaux 1, CNRS/IN2p3, Centre d'Etudes Nucl\'eaires de Bordeaux Gradignan, 33175 Gradignan, France}
\email{rwr@astro.stanford.edu}
\altaffiltext{1}{Einstein Fellow}

\begin{abstract}

Most pulsars observed by the {\it Fermi} LAT have $\gamma$-ray
luminosities scaling with spindown power ${\dot E}$ as
$L_\gamma \approx ({\dot E}\, \cdot \, 10^{33}{\rm erg/s})^{1/2}$.
However, there exist one detection and several upper limits an order
of magnitude or more fainter than this trend. We describe these
`sub-luminous' $\gamma$-ray pulsars, and discuss the case for this
being an orientation effect. Of the 12 known young radio pulsars 
with ${\dot E}>10^{34} {\rm erg\,s^{-1}}$ and $d\le 2$\,kpc 
several are substantially sub-luminous. The limited available 
geometrical constraints favor aligned geometries for these pulsars,
although no one case for alignment is compelling. In this scenario GeV 
emission detected from such sub-luminous pulsars can be due to a 
lower altitude, lower-power accelerator gap.
\end{abstract}


\section{Introduction}

	The Large Area Telescope (LAT) on the {\it Fermi} satellite has
now detected over 75 spin-powered pulsars \citep{psrcat,srdrev}. Among the 
$\approx 50$ non-recycled energetic pulsars there is a clear trend for
$\gamma$-ray `efficiency' to increase with decreasing spin-down power
${\dot E}$, giving a heuristic $\gamma$-ray luminosity
$$
L_{\gamma,heu} \approx ({\dot E}\, \times \, 10^{33}{\rm erg/s})^{1/2}.
\eqno(1)
$$
This is a natural result for models where the emission is produced
by a Goldreich-Julian current of charges passing through a characteristic 
potential drop \citep{ha81,a06}.
Of course, energy conservation limits $L_\gamma < {\dot E}$, and as ${\dot E}$
decreases, the star is unable to maintain the potential drop, leading to a
`death zone' below ${\dot E} \approx 10^{33}-10^{34}{\rm erg\, s^{-1}}$ where 
this process starts to turn off. This is portrayed in figure 5 of \citet{psrcat},
where most energetic pulsars lie between Eq (1) and unit efficiency. Only two
young pulsars in that plot lie significantly below the $L_{\gamma,heu}$ line: PSR J0205+6449,
where a small inferred distance places it just below this value, and PSR J0659+1414
(to be discussed in this paper) which is $\sim 20\times$ less luminous. Thus,
independent of its physical validity, Eq. (1) forms an effective lower
luminosity envelope to the bulk of the observed pulsar sample.

	Estimates of $L_{\gamma}$ suffer two complications.
The first is the source distance; for most LAT pulsars we have only
distance estimates based on the pulsar Dispersion Measure (DM).
DM modeling \citep[hereafter CL02]{cl02} is believed to provide statistically useful estimates
of pulsar distances, with a scatter of $\approx 30$\% about independent distance
estimates, although typical errors for nearby pulsars may be as large as 60\% 
(see Deller 2009). DM distances are certainly not reliable for individual objects,
and it appears \citep{psrcat} that they may be especially poor for the
young, energetic LAT pulsars. This is likely since the sample is nearby and associated
with regions of active star formation where the excess ionized gas may significantly perturb
the dispersion measures. About a third of the LAT pulsars are found directly in
the $\gamma$-ray data through so-called `blind' searches \citep{blind,sp10}; 
most of these lack
radio detections and so do not even have DM distance estimates.
The second complication is the conversion from the observed energy flux
$F_{\rm E}$ along the Earth line-of-sight to the true sky averaged luminosity
$$
L_\gamma = 4\pi f_\Omega F_{\rm E} D^2.
\eqno(2)
$$
\citet{wet09} and \citet[RW10]{rw10} have estimated `flux conversion factors'
$f_\Omega$ for this correction for a variety of pulsar models and viewing geometries.
For most of the observed pulsars, $f_\Omega$ should be in the range $0.7-1.3$,
although some lower ${\dot E}$ pulsars, especially $\gamma$-selected objects
\citep{wr11}, may have $f_\Omega$ as small as 0.1 for `outer gap' (OG) geometries.

\begin{figure*}[t!!]
\vskip 8.5truecm
\includegraphics{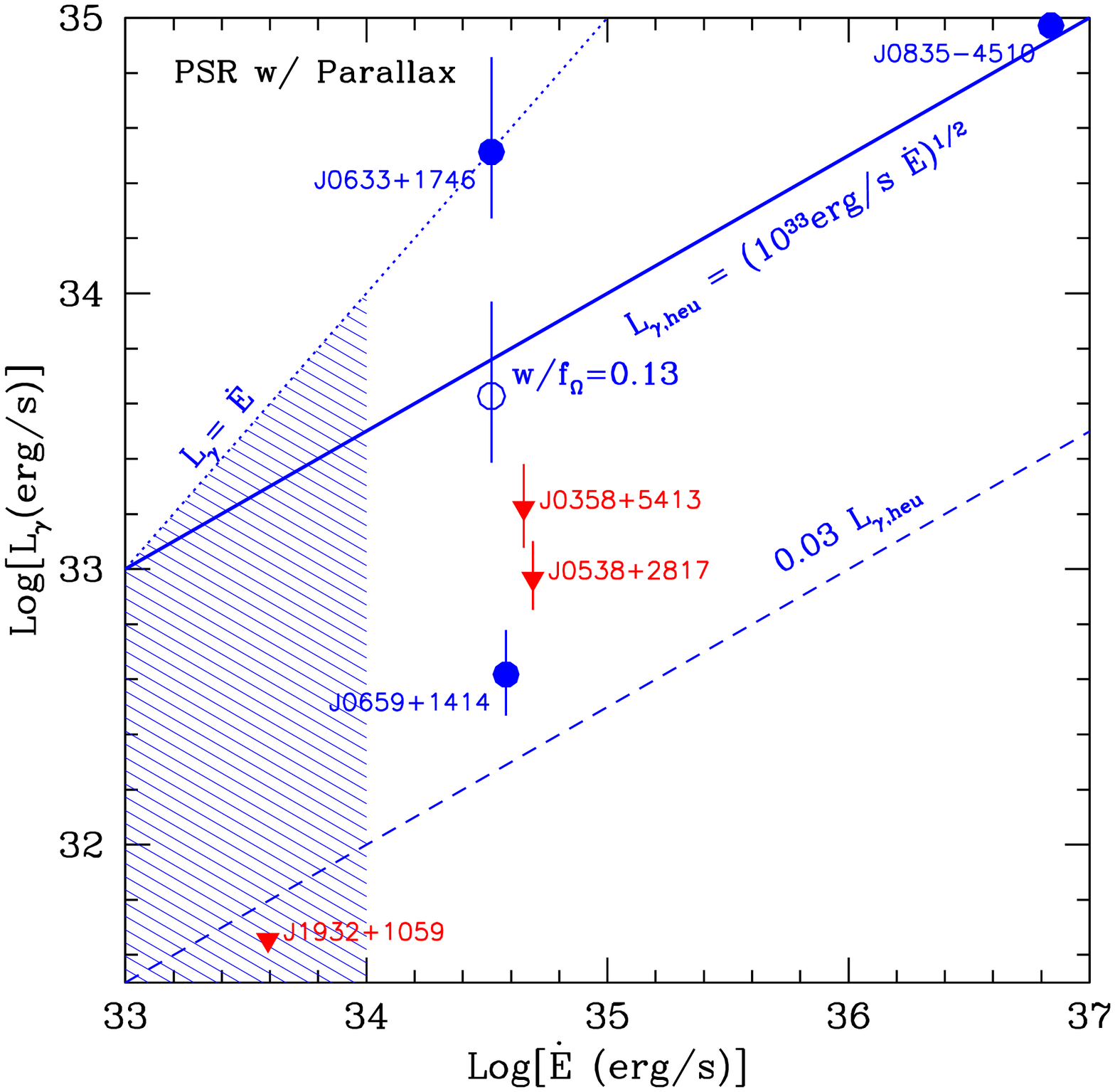}
\includegraphics{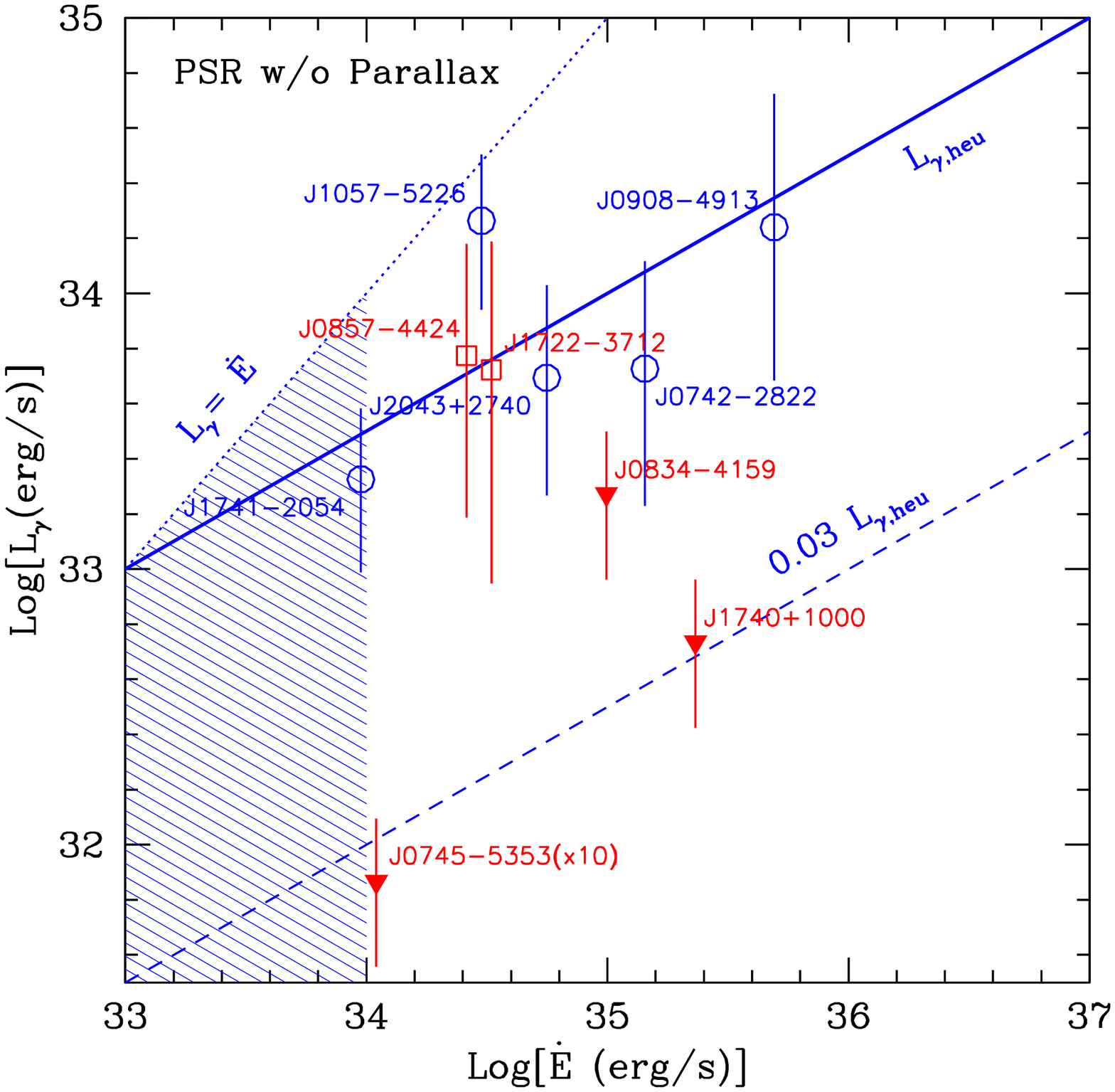}
\begin{center}
\caption{\label{LgamEdot} The spin-down-luminosity plane for energetic pulsars,
with the heuristic luminosity trend, which saturates somewhere in the 
`death zone' (shaded).
Unpulsed (DC) $E>0.1$\,GeV luminosities or limits are plotted, assuming 
$f_\Omega=1$ (the RW10 $f_\Omega=0.13$ point for Geminga is also shown). 
Left: Objects with parallax distance measurements.  The 95\% error bars for LAT-detected 
objects (circles) include the flux imprecision, but are dominated by
the parallax uncertainty. For the 95\% upper limits (triangles), the error flags
represent the parallax uncertainty.
Right: Objects with DM-estimated distances. 
Circles: The DC luminosities for radio pulsars with LAT pulse detections,
Squares: LAT DC detections (this paper), Triangles: DC upper limits. All error bars
include an assumed 30\% DM distance uncertainty. For a few of the fainter
LAT detections flux uncertainties contribute significantly.
For PSR J0745$-$5353 the luminosity at the DM-estimated distance is 
$10\times$ lower than the point shown.
}
\end{center}
\end{figure*}

	However, there are a handful of pulsars whose observed luminosity or limit
fall an order of magnitude or more {\it below} $L_{\gamma,heu}$.
In spite of the uncertainties just discussed we can make a case that
they are truly sub-luminous. There are three possible interpretations. The first
is that the $\gamma$-ray radiation is beamed away from the Earth line-of-sight
(or equivalently $f_\Omega > 10$).
The second is that some particular physical property of the pulsar prevents
them from producing the bright high-altitude $\gamma$-ray emission typical
of other energetic pulsars. The third is that the DM distance is especially
poor and the pulsar is much more distant than estimated. We test here the first
possibility, that $\gamma$-ray beaming explains the low observed fluxes of
some nearby energetic pulsars. We also comment briefly on the possibility
that objects with detected luminosities $\ll L_{\gamma,heu}$ may be probing an emission 
component different to the powerful high-altitude gap emission which apparently 
dominates the bulk of the LAT-detected pulsars.

\section{The Sub-Luminous Pulsar Candidates}

To find sub-luminous pulsars, we measure the DC (unpulsed) flux at the
positions of nearby ($d \le 2$ kpc), energetic ($\dot{E}>10^{34}\,\mathrm{erg\,s^{-1}}$)
non-recycled radio pulsars selected from the ATNF pulsar catalog \citep{met05}. 
There are 12 such objects (Table 1 also includes two comparison objects).  
Since the LAT has detected
several pulsars, especially millisecond pulsars, in the $\sim 10^{33-34}{\rm erg\,s^{-1}}$
boundary of the `death zone' we also consider the well-studied nearby
${\dot E}  = 10^{33.6}{\rm erg\, s^{-1}}$ pulsar PSR J1932+1059 (B1929+10), which has a 
low LAT flux limit. Finally, for comparison we include Geminga (J0633+1746), 
a nearby $\gamma$-selected pulsar
with an HST parallax measurement. We should note that this distance cut-off
is somewhat arbitrary; for example PSR J1747$-$2958 with a CL02 distance of
2.01\,kpc is a LAT pulsed detection.

To measure the unpulsed fluxes, we use 24 months of LAT data 
(Aug 4 2008 -- Aug 4 2010) and the P6\_V11 instrument response function, 
a refinement to previous analyses reflecting improved understanding of the point spread function
and effective area \citep{p6v11}. `Diffuse-Class' events were selected from good runs
with rocking angle $<52^\circ$, reconstructed energies 
$-0.75 < {\rm Log} (E_\gamma/{\rm GeV}) < 2$, and a reconstructed zenith 
angle $<100^{\circ}$. The list of point sources used in the background model 
is drawn from a preliminary version of the two-year {\it Fermi} catalog.
The analysis used an updated version of the model for the diffuse background 
- Galactic, extragalactic, and residual cosmic rays - that is being prepared 
for publication by the LAT team. Like the model used for the 1FGL catalog \citep{1FGL}
it is based on fitting templates for the diffuse emission to the LAT data.

For each pulsar we assume an exponentially cutoff spectrum $dN/dE =
N_0\,(E/\mathrm{GeV})^{-\Gamma}\,\exp(-E/E_c)$. For the bright LAT-detected pulsars
(marked $^b$ in the Table) we allow $E_c$ and $\Gamma$ to vary in the fits; the 
results are consistent with parameters quoted in \citet{psrcat}. For the other pulsars 
we set these parameters to values determined from an empirical fit to detected LAT 
pulsars (RW10): $\Gamma=-4.1 + 0.156\log_{10}\dot{E}$ and
$E_c/\mathrm{GeV}=-0.45+0.71\log_{10}B_{LC}$, with $B_{LC}$ the magnetic field
measured at the pulsar's light cylinder. We evaluate the likelihood for $N_0$ at the
known pulsar position using `pointlike', a binned likelihood analysis tool \citep{kerr10},
and using a Bayesian approach with a uniform prior we integrate the likelihood to
$97.5\%$ to obtain a $2\sigma$ upper limit on the flux. 
For sources with apparent DC emission, we determine the corresponding 95\% 
range for the measured $N_0$. For comparison with results in \citet{psrcat}
these measurements and upper limits are then converted to $E>0.1$\,GeV fluxes using the model spectra.
The uncertainties reported for the measured fluxes are statistical only; additional 
systematic error arises from uncertainty in the effective area of the LAT (about 20\% 
below 1 GeV, 10\% at 1--10 GeV, and 30\% above 10 GeV) and the structure of the
diffuse background.  Systematic uncertainties in the upper limits stem
primarily from uncertainty in the background model and are comparable
in magnitude to those associated with the assumed beaming factor discussed below.

\begin{deluxetable*}{lcccccccccc}[t!!]
\tablecaption{\label{PulsarParameters} Young Local, Energetic Radio Pulsars: DC fluxes and Geometry Constraints}
\tablehead{
  \colhead{Name} & \colhead{log(${\dot E}$)} & \colhead{log($\tau_c$)} & \colhead{$d\tablenotemark{c}$} &$F_{{\rm E},>0.1{\rm GeV}}$
    & \colhead{$W_{10}$} & \colhead{$W_{1}$}& \colhead{$\Delta\phi_{PA}$} & $h_{LC}$ & $B_{LC}$  & Ref.\cr
 & [${\rm erg\,s^{-1}}$] & [y] & [kpc] &  [$10^{-12}{\rm erg\,cm^2\,s^{-1}}$] &[deg] &[deg] & [deg]$^d$& &[kG] & $^e$
}
\startdata
{\bf J0358+5413}   &34.65&5.75&$1.04^{+0.21}_{-0.16}$ &$<12.8$               &38.9&56&13.5&0.059& 2.1& 1\\ 
J0534+2200$^b$   &38.66&2.98$^f$&2.0                    &$1828.^{+3.}_{-21.}$  &   &   &   &      & 980.& \\ 
{\bf J0538+2817}  &34.69&4.60$^f$&$1.30^{+0.22}_{-0.16}$ &$<4.5$                &55.4&92&35.0&0.153& 2.3& 2 \\  
J0633+1746$^{a,b}$&34.52&5.53 &$0.25^{+0.12}_{-0.06}$ &$4340.^{+34.}_{-29.}$ &--  & --&-- & --  & 1.2&  \\
{\bf J0659+1414}$^b$  &34.58&5.05&$0.288^{+0.033}_{-0.027}$&$41.6^{+6.8}_{-5.8}$&31.3&54&13.9&0.061& 0.8& 3\\ 
{\bf J0745$-$5353} &34.04&6.10&0.25                     &$<0.98$              &34.9&61& 4.0&0.017& 0.8& unpub.\\
J0834$-$4159 &34.99&5.64&1.66                     &$<5.6$              &24.0&  &    &     & 3.9&  \\ %
J0835$-$4510$^b$&36.84&4.05&$0.287^{+0.019}_{-0.017}$&$9466.^{+3.}_{-3.}$ &16.9&27& 4.2&0.018& 45.& 4\\   
J0857$-$4424 &34.42&5.35&1.94                     &$13.1^{+6.6}_{-6.2}$&18.7&  &    &     & 0.8&   \\  
J1057$-$5226$^b$&34.48&5.73&0.72                    &$294.^{+9.5}_{-8.2}$ &31.0&  &    &     & 1.3&   \\
J1722$-$3712 &34.52&5.53&1.85                    &$12.8^{+10.0}_{-8.5}$&13.3&22&12.0&0.052& 1.2& unpub.\\ 
{\bf J1740+1000}  &35.36&5.06&1.24                    &$<2.9$               &42.5&71&28.0&0.122& 4.7& unpub.\\
{\bf J1932+1059}$^a$&33.59&6.49&$0.361^{+0.010}_{-0.009}$&$<2.9$              &19.1&55& 4.5&0.020& 0.4 & 4 \\	
J2043+2740$^b$  &34.75&4.08$^f$&1.80                     &$12.7^{+3.6}_{-3.0}$&16.6&32& 8.0&0.035& 3.7& 5 \\  

\enddata
\tablenotetext{a}{Comparison pulsars, not members of the uniform radio-loud, ${\dot E}>10^{34}{\rm erg\, s^{-1}}$, $d\le2$\,kpc set.}
\tablenotetext{b}{{\it Fermi} LAT pulsed detection.} 
\tablenotetext{c}{Parallax distances from the ATNF data base. CL02 DM distances (w/o errors). Classical Crab kinematic distance.}
\tablenotetext{d}{$\Delta\phi_{PA}$ measured from the $W_{10}$ pulse center.}
\tablenotetext{e}{Reference for the polarization profile used to fit the pulse widths, offsets and 
polarization sweeps: 1=\citet{gl98}, 2=\citet{vhx97}, 3=\citet{ew01}, 4=\citet{jhv05}, 5=\citet{net11}}
\tablenotetext{f}{Age of the associated supernova remnant; $\tau_c$ is larger, implying initial spin period $P_0 \sim P$.}
\end{deluxetable*}

	Our study gives six candidate sub-luminous pulsars (marked in {\bf bold}). 
Three in the uniform sample (plus PSR J1932+1059 = B1929+10) have parallax
distance measurements. These are particularly important as the parallax constraints
control a major factor in the luminosity uncertainty, allowing us to probe
the effects of beaming geometry and gap emissivity.
For the others we must rely at present on the
DM distance estimates. These pulsars are displayed in Table 1 and Figure 1.
Figure 1 also shows several other nearby non-recycled LAT-detected pulsars, highlighting
the separation of our sub-luminous set from this sample. For this figure
we have assumed $f_\Omega=1$ for all pulsars. The plotted luminosity 
errors are dominated by the distance uncertainties, but do include the statistical
flux errors. Of course, systematic errors and non-unity $f_\Omega$ may add 
additional uncertainty for individual pulsars.

\section{External Angle Constraints}

	For simple dipole models (e.g. the OG model) the pulse
profile and the expected radiation on the Earth line-of-sight are 
determined by the magnetic inclination angle $\alpha$ and the viewing 
angle $\zeta$. If these angles are known, we can predict $\gamma$-ray
pulse profiles and fluxes for specific models and correct observations to
the true $L_\gamma$. Unfortunately these are poorly known in many cases.

\subsection{Radio Polarization Data}

The sub-luminous candidates treated here are known radio pulsars, so
the magnetic impact angle $\beta=\zeta-\alpha$ is
believed to be small. In the context of the rotating vector model
\citep{rc69} radio polarization data can constrain the viewing angles.
In most cases, the small range of phase illuminated by the radio
pulse allows only an estimate of the magnetic impact angle
$$
\qquad\quad\beta = \zeta - \alpha \approx 
{\rm sin^{-1}} [{\rm sin} \alpha/({\rm d}\Psi/{\rm d}\phi)_{\rm max}]
\eqno (3)
$$
where the maximum rate of the polarization position angle (PA) sweep $\Psi(\phi)$
occurs at $\phi_{\rm d\psi, max}$, near the closest approach to the magnetic axis. Here the sign
of the sweep is meaningful, determining whether the line of sight is closer to
or farther from the positive rotation axis than the observed magnetic pole (at
inclination $\alpha$). Occasionally, when the radio pulse is very broad or when the pulse profile
presents an inter-pulse, the radio polarization can make meaningful estimates
of both $\alpha$ and $\zeta$, from fits to the full polarization sweep
$$
{\rm tan}(\Psi + \Psi_0) = 
{{{\rm sin}\alpha\, {\rm sin} (\phi-\phi_0)} \over 
{{\rm sin}\zeta\,{\rm cos}\alpha - 
{\rm cos}\zeta \,{\rm sin}\alpha\,{\rm cos}(\phi-\phi_0)}
}
\eqno (4)
$$
where the polarization has the absolute position angle $\Psi_0$ at $\phi_0$.
\citet{keith10} have recently presented several examples of constraining fits 
of Eq. (4) to high quality polarization data. As described by \citet{ew01},
while nearly all authors fit to Eqs. (3) and (4), given the standard astronomical
convention of position angle measurement (increasing N through E) these 
equations are inconsistent with pulsar angles increasing from the positive spin 
axis (the `RVM convention problem'). 
To be consistent, one must actually use $\alpha_{EW01} = \pi-\alpha_{RVM}$
and $\beta_{EW01} = -\beta_{RVM}$. Usually this correction is only a formality,
but as fits to the $\gamma$-ray emission improve, including details of sweep-back
and magnetospheric currents, the signs can be important. Thus in the figures
and discussion to follow, we convert all `RVM'-fit angles to the consistent \citet{ew01}
convention; we encourage future workers to do the same.

	Other phenomenological constraints may be extracted from the radio
data. For example, radio emission is generally believed to be produced within
the `open zone' above the polar cap. For a static aligned dipole the half opening
angle covered by this radio beam is
$$
\rho = 3/2 h_{LC}^{1/2}
\eqno (5)
$$
radians for modest emission altitudes $h_{LC} = 2\pi h/Pc$. If the observed radio pulse fills this
cone we can write $h_{LC}$ in terms of the pulse width $W \approx 2\rho$ 
$$
h_{LC} = 4/9\, {\rm Acos}^2 [{\rm cos}\alpha\,{\rm cos}\zeta + 
   {\rm sin}\alpha\,{\rm sin}\zeta\,{\rm cos}(W/2)];
\eqno (6)
$$
if the radio emission does not fill the open zone this provides a lower limit
for the emission height.
It has also been shown that, due to a combination of field line sweep back and aberration,
the phase of the center of the radio pulse $\phi_I$ should lead the phase of the
max PA sweep rate by 
$$
\Delta \phi_{\rm PA} \approx 4h_{LC}
\eqno (7)
$$ \citep[eg.]{bcw91,d08}. Observationally we identify $\phi_I$ with the mid-point of the
pulse at 10\% of its peak and $\phi_{\rm PA}$ is identified with $\phi_0$ in an RVM fit. 
The true phase of minimum magnetic angle is between  $\phi_I$ and $\phi_{\rm d\psi, max}$.
These expressions assume simple static dipoles and low altitudes. We have checked
against detailed numerical simulations of swept-back dipole magnetospheres
and find that the actual pulse intensity center and phase of maximum PA sweep are both
sensitive to details of the magnetic field structure, especially conditions at the light cylinder
that define the edge of the open zone \citep{crj11}. These differences are modest
at $h_{LC} < 0.05$. For objects indicating higher altitude radio emission 
detailed comparison with the numerical results can be important.

	In practice, radio pulse profiles may represent
`patchy' illumination of the radio zone \citep{lm88},  even for these young pulsars. This complicates
our estimates of $W$ and $\phi_I$. In Table 1, we list both $W_{10}$, the full width
of the radio pulse at 10\% of the peak intensity and $W_1$, an estimate of the
pulse width at 1\% of the peak.  These measurements were made on archival 1.4\,GHz
profiles (see Table references).
The $W_1$ estimate is necessarily approximate,
especially for the lower S/N pulse profiles. At such low flux levels, extended pulsed emission
may be generated by interstellar scattering tails, weak emission components unassociated
with the main dipole cap or even non-linearities in the measurement system. 
Nevertheless, for at least a few of these pulsars, this broader width captures 
weak components of the pulse coming from the principal emission zone.  
Further, in some cases, the assumption of pure dipole geometry
and even the identification of the radio beam with the open zone are suspect. 
However, despite all of these caveats, these radio measurements do provide some 
phenomenological constraints on the range of allowable $\alpha$ and $\zeta$,
even when values for individual pulsars are suspect.

\subsection{PWN Torus fits}

	When the pulsar wind momentum is equatorially concentrated, one
may observe a `torus' of emission at the spin equator, as for the Crab
and Vela pulsars. The most useful examples are found in {\it Chandra} 
(CXO) X-ray images of PWN tori, where synchrotron emission is produced in
the mildly relativistic flow downstream from the termination shock. Doppler
boosting allows one to distinguish the `front' and `back' sides and so
fitting can measure the spin axis inclination to the line of sight
\citep{nr08}. The images are not, however, sensitive to the sign of the
spin, so a torus fit cannot distinguish between $\zeta$ and 
$\zeta^\prime = \pi-\zeta$. Occasionally, symmetric jets also allow $\zeta$ 
estimates. These fits provide relatively robust
model-independent constraints on $\zeta$, largely orthogonal to the RVM
measurements, so that the combination provides a good picture of the pulsar geometry.

	Unfortunately, to date no strongly `sub-luminous'
pulsar has an X-ray torus measurement, since these are typically available only for
very young $\tau < 10^{4.5}$y pulsars. However, PSR J1930+1852 (unseen by the LAT) has a fit
angle $\zeta = 33\pm3^\circ$ suggesting that its OG emission should not be visible
\citep{nr08}; as the LAT exposure increases this can eventually be a useful
comparison. Also there is some hope of obtaining $\zeta$ estimates for older
(even millisecond) pulsars from fits to the geometry of H$\alpha$ bow shocks; in some
cases \citep[eg.][J1741$-$2054]{ret10} these show clear signs of equatorially concentrated momentum
flux and thus opportunities to constrain $\zeta$.

\begin{figure}[h!!]
\vskip 7.9truecm
\includegraphics{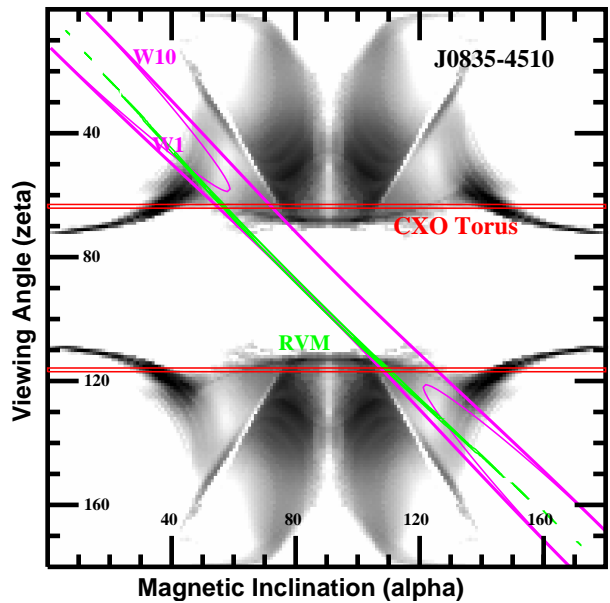}
\begin{center}
\caption{\label{VelaEx} The spin geometry plane for Vela, with various observational constraints.
The background gray scale shows the goodness of fit of the LAT light curve to a
basic OG model, with dark colors better agreement.
}
\end{center}
\end{figure}

\begin{figure*}[hb!!]
\vskip 15.9truecm
\includegraphics{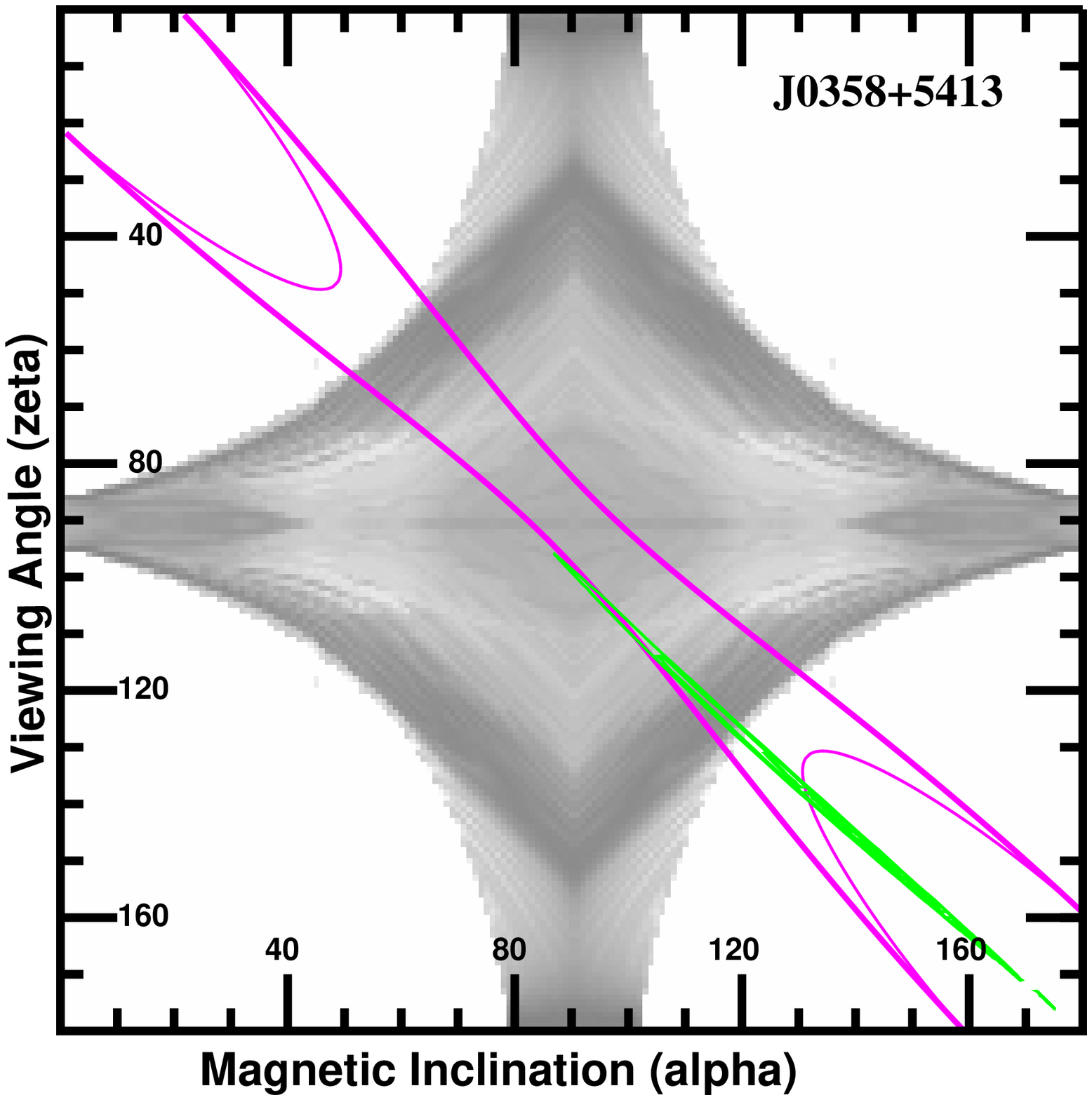}
\includegraphics{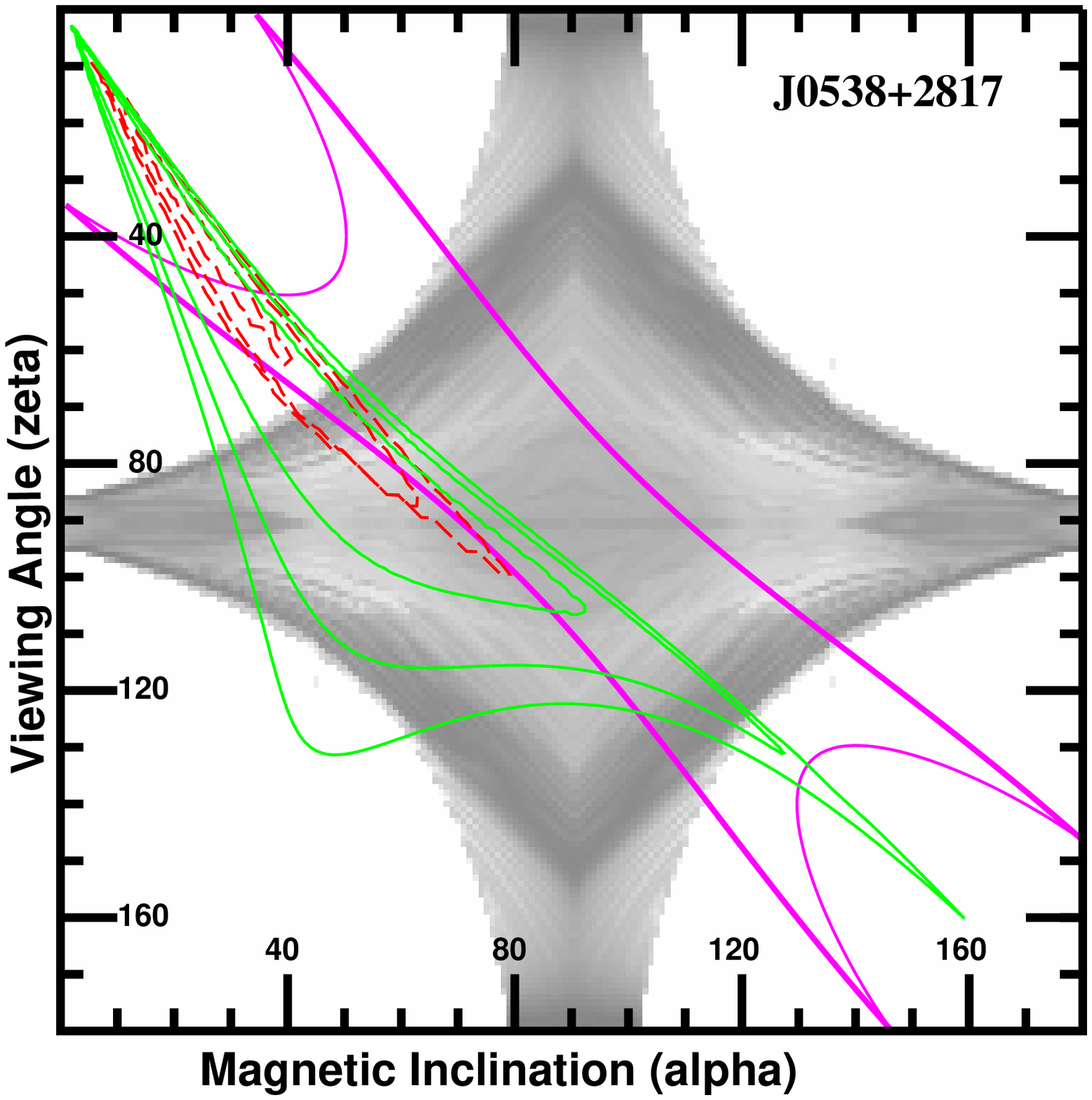}
\includegraphics{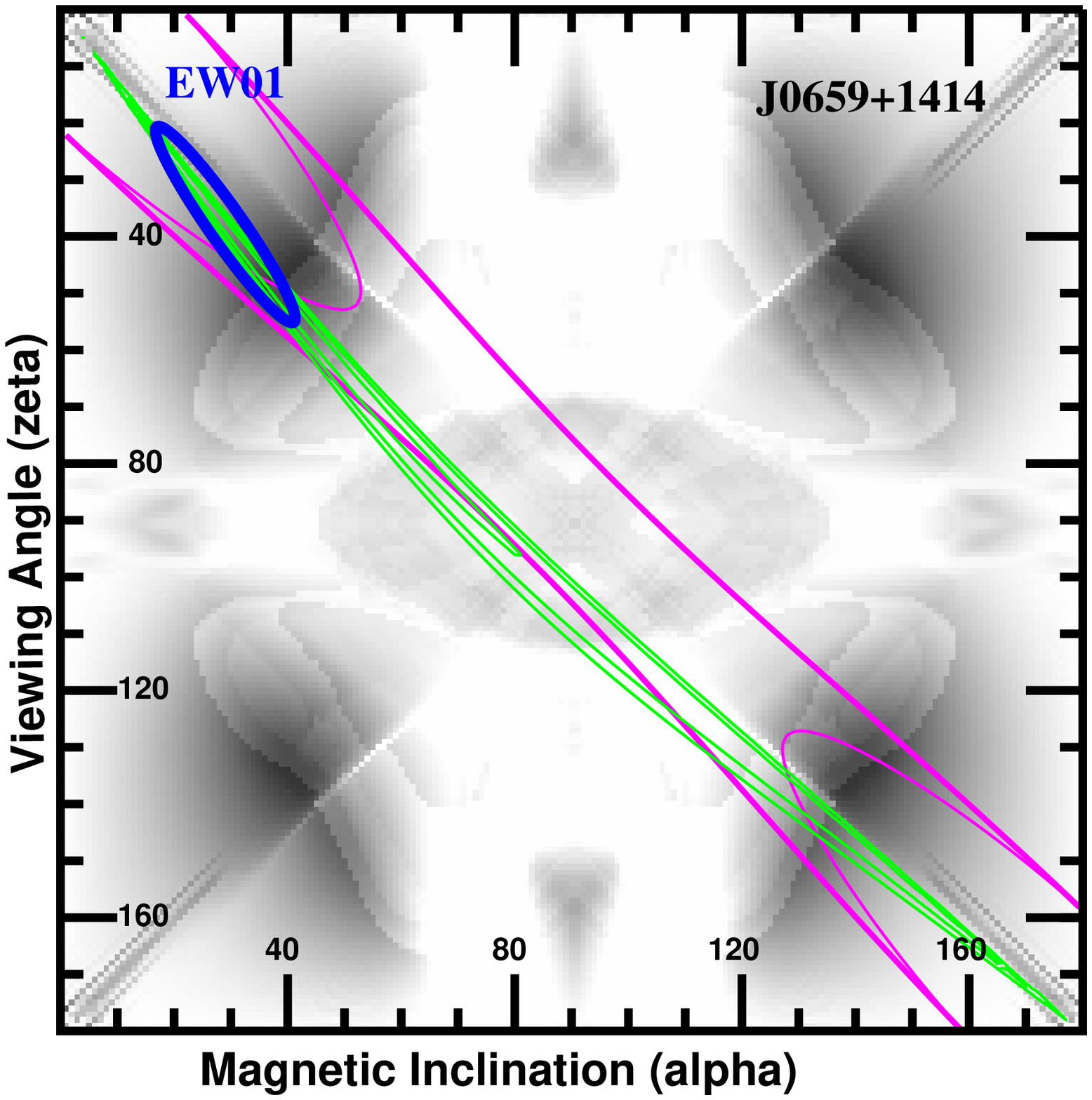}
\includegraphics{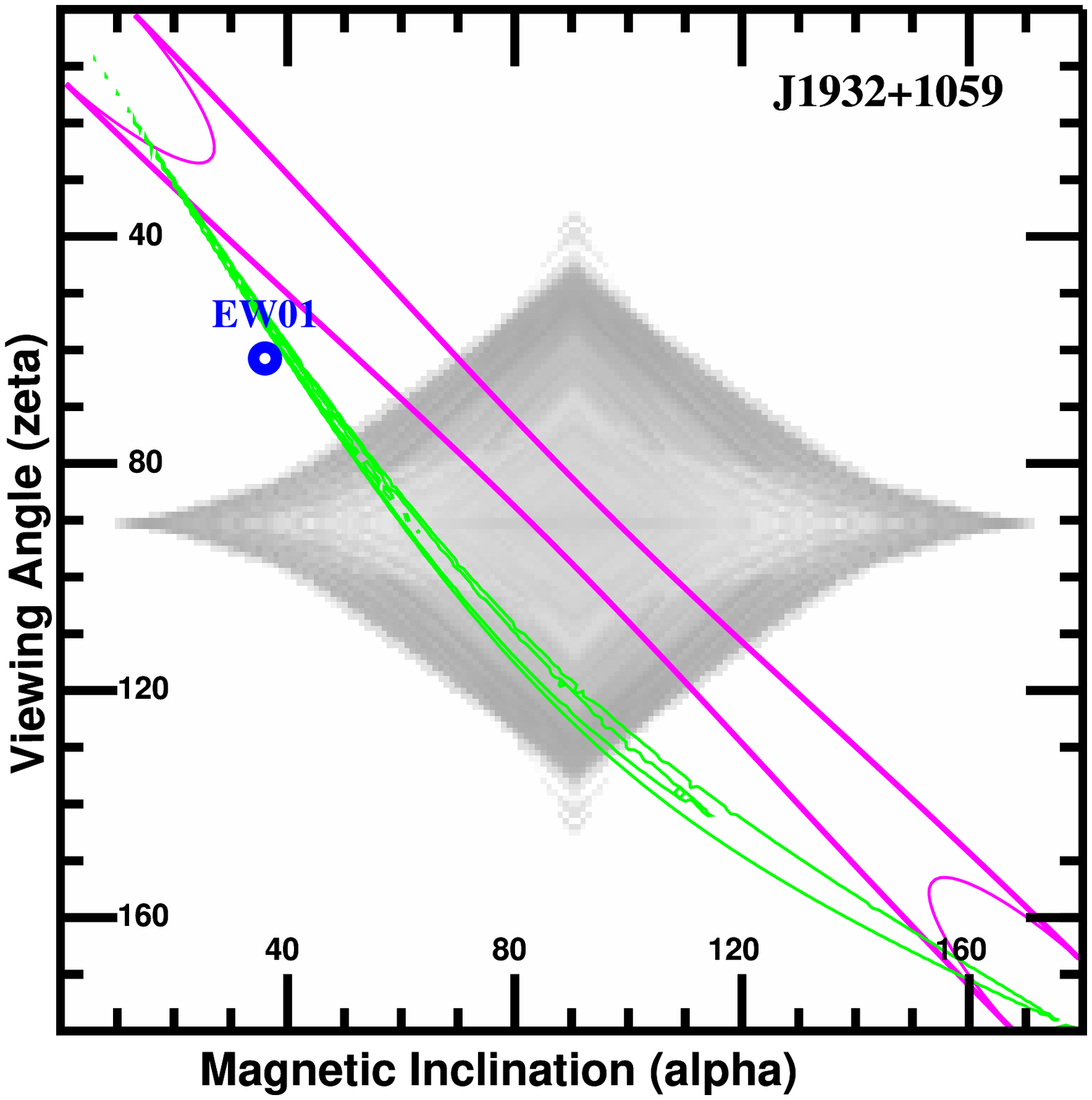}
\begin{center}
\caption{\label{plx_const} The spin geometry plane for sub-luminous pulsars with 
parallax distances. The backgrounds show the generic locations providing
sharp OG pulses, except for PSR J0659+1414, where the background shows the
allowed fits to the observed LAT pulses, including lower altitude 
(eg. `Two Pole Caustic' TPC)
emission. Three green contours show the loci of best RVM matches, while the
bold and narrow magenta curves showed the regions allowed by emission from
the static dipole open zone for our estimated $h_{LC}$. For PSRs J0659+1414 (B0656+14)
and J1932+1059 (B1929+10) the RVM fits of \citet{ew01} are indicated.
For PSR J0538+2817 the fits imply large $h_{LC}$, requiring
a numerical magnetosphere model. The fits to the polarization geometry using
such models are shown by the dashed (red) contours.
}
\end{center}
\end{figure*}

\section{Individual Objects}

	We can combine these various geometry constraints to restrict the viable
location of a pulsar in the ($\alpha,\,\zeta$) plane and for this location compare with
the predictions of the various $\gamma$-ray emission models. As an example,
we show in Figure 2 the constraints for the bright, well-studied
Vela pulsar PSR J0835$-$4510 (B0833$-$45). The RVM model was originally developed for Vela \citep{rc69}
so it is not surprising that the high S/N Vela data provides good constraints on $\beta$.
As usual $\alpha$ is not well determined. However, there is a good measurement
of $\zeta$ from {\it CXO} images of the X-ray torus \citep{nr08}.  Finally the pulse width
constraints are shown. The background gray-scale shows the `goodness of fit' of 
the LAT Vela profile to an Outer Gap light curve computed for the particular $\alpha$
and $\zeta$, assuming a current-free retarded dipole field structure (RW10).

	The contours marked RVM are quite crowded, since the high S/N Vela
data provide very strong $\beta$ constraints. These cross the $\zeta$ constraints
from the X-ray torus and the two constraints select solutions at 
$(\alpha,\, \zeta) = (56^\circ,\,63.5^\circ)$ and $(109^\circ,\,116.5^\circ)$.
For the estimated $\Delta\phi_{PA}$ and the resulting $h_{LC} \approx 0.018$,
the pulse width $W_{10}$ is easily consistent with these solutions. They
are not consistent with the $W_1$ width, but such a wider pulse could be easily
accommodated if this faint emission comes from slightly higher altitudes.
Indeed, evidence of variable pulse components and micro-structure in the Vela
pulse wings \citep{jet01} suggests such multi-altitude emission. The darker gray
scales in the background show regions with good fits to the LAT light curve.
Evidently, the best regions in this model are not at the RVM+PWN preferred positions.
As shown in RW10, model perturbations such as magnetospheric currents can
shift the locations of the best fits; for example OG models with currents can 
produce reasonable agreement with the $(109^\circ,\,116.5^\circ)$ solution. 
However other models remain viable. For example the `Separatrix Layer' (SL) model
of \citet{bs10} produces a good match at the $(56^\circ,\,63.5^\circ)$ solution.

\subsection{ Pulsars with Parallaxes}

In Figure \ref{plx_const} we show the constraints for the sub-luminous candidates
having parallax measurements.  The background gray scales show the region 
where a narrow OG, with gap widths $w \approx L_{\gamma,heu}/{\dot E}$, 
produces sharp caustic pulses. The gray levels indicate goodness of fit for a generic single
$\gamma$-ray pulse. Of course with an actual LAT detection the detailed light
curve and phase produce much more detailed constraints within this envelope
(see Figure \ref{VelaEx}). Note that sharp OG pulses are not expected
(white background) near the spin poles unless the pulsar is an orthogonal rotator.
In contrast, lower altitude (e.g. TPC) models produce emission at small
$\beta$ all the way to the poles and good pulse matches are expected in the
OG blank zones, if such low altitude gaps are active.

{\bf PSR J0358+5413} (B0355+54) is a bright radio pulsar
for which RVM fits provide good $\beta$ constraints and a preference for 
$\alpha > 110^\circ$. At the estimated $\Delta\phi_{PA}$, the $W_{10}$ 
pulse width does not significantly tighten this bound, but if 
we include the wider $W_1$ width the constraints tighten with $\alpha > 130^\circ$
and $\zeta > 140^\circ$. This is also the region of best RVM fits.
Thus while the best-fit geometry localizes to $(\alpha,\, \zeta)$
where OG emission would not be visible, acceptable solutions include
the range that could be consistent with visible high altitude $\gamma$-rays. 
Since our present flux bound only restricts us to $<L_{\gamma,heu}/3$ we 
regard this as a plausible sub-luminous pulsar, but not a strong case. Increased
LAT exposure and improved geometrical constraints are needed to make this
definitive.

\begin{figure*}[t!!]
\vskip 9.3truecm
\includegraphics{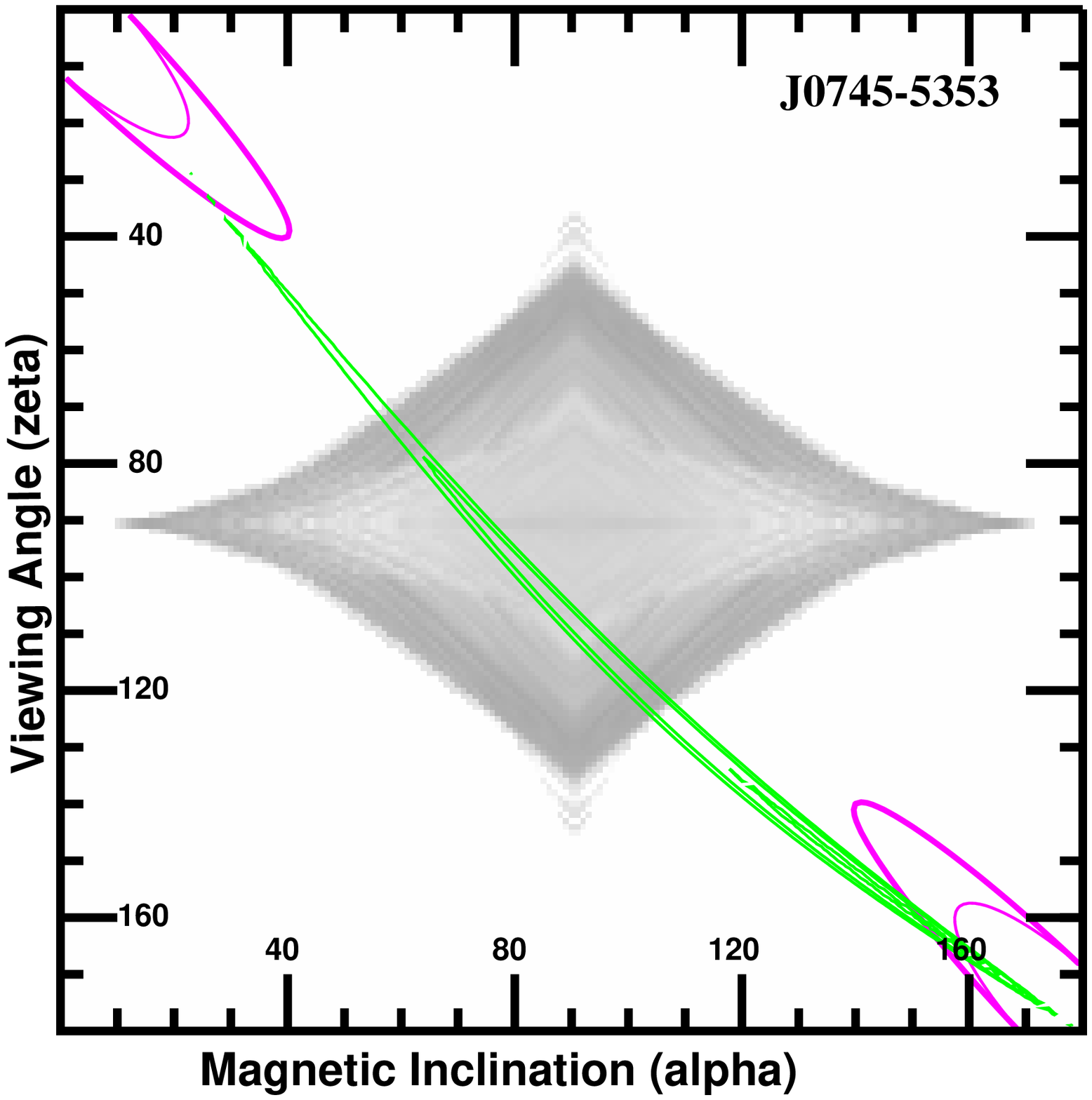}
\includegraphics{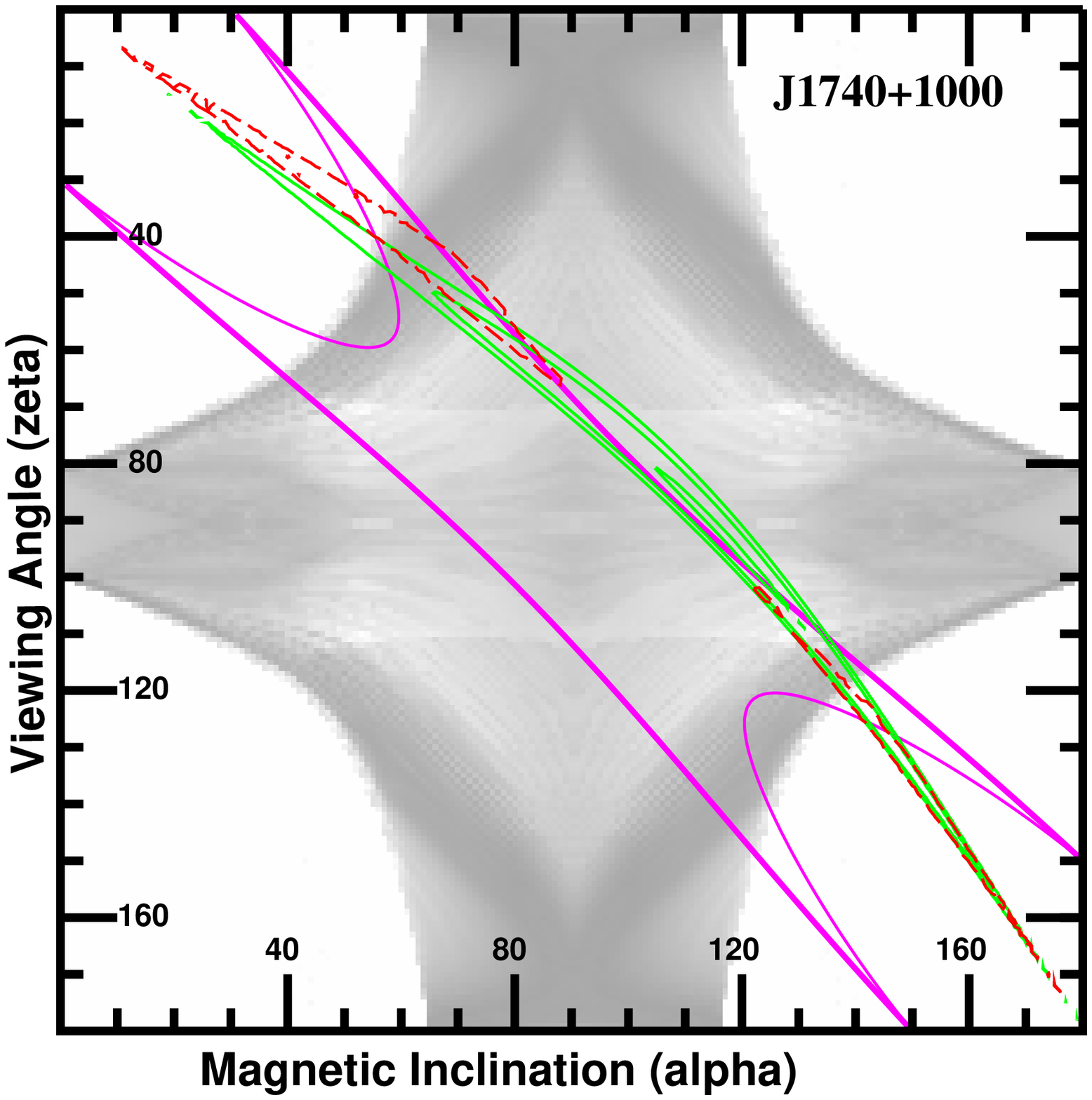}
\begin{center}
\caption{\label{noplx_const} The spin geometry plane for two pulsars without parallax distances,
showing the radio polarization and pulse width constraints. For geometries
away from the gray background, the sources are not expected to have strong 
outer magnetosphere $\gamma$-ray pulses. The PSR J1740+1000 data suggest large
$h_{LC}$ requiring numerical modeling; the locus of best fits for these models 
is shown by the dashed (red) contours.
}
\end{center}
\end{figure*}

{\bf PSR J0538+2817} in contrast has only rather poor RVM constraints. However, the
large pulse width does help restrict the range of viable solutions, even
though $h_{\rm LC}$ is high. Here the best combined radio data seem to prefer
small $\alpha$, small $\zeta$. In fact if the $W_1$ constraint is included
we conclude $\alpha < 35^\circ$, $\zeta< 50^\circ$ so we would not expect 
to see OG emission. However, using only the $W_{10}$ width a
wide range of $\alpha$ is allowed towards the edges of the RVM and pulse width
contours. {\it CXO} in fact shows a small X-ray PWN around the pulsar \citep{rn03,net07}.
Unfortunately, the emission is too faint and compact to provide a good $\zeta$ measurement,
although the existence of opposing jet-like features suggest
$\zeta \approx 90^\circ$. Interestingly, \citet{ket03} find an RVM fit 
giving $\alpha \approx 85^\circ$ and $\beta \approx -2^\circ$, which would be consistent
with jets viewed near side-on; however this solution is well outside the
RVM-allowed region in Figure \ref{plx_const}. 
It is worth noting that with this pulsar's large $h_{LC}=0.14$
the detailed field treatment of \citet{crj11} can be important.
The dashed (red) contours show the constraints from fits to these numerical
models. The best fit altitude depends on $\alpha$ and $\zeta$, ranging from
$h_{\rm LC}=0.13 - 0.17$. The contours show best solutions to these numerical models,
marginalized over $h_{LC}$. Good solutions are found for $\alpha<40^\circ$,
acceptable solutions are at $\alpha<80^\circ$.
Again we must conclude that alignment is preferred,
but nearly orthogonal rotators are not excluded. This is a case where
improved radio observations with higher S/N can substantially
improve the polarization modeling constraints.

	The next object, {\bf PSR J0659+1414} (B0656+14) is the archetype sub-luminous pulsar.
This object was included in the very careful polarization study of \citet{ew01}, who find
$\alpha = 29 \pm 23^\circ$ and $\beta=8.9\pm 6.1^\circ$ (blue ellipse). 
Earlier, \citet{lm88} found
$\alpha \approx 8.2^\circ$ and $\beta\approx 8.2^\circ$, so there has been some
consensus that this is an aligned pulsar. However, we and \citet{wel10} 
find less strong constraints on $\alpha$ even though we are fitting the same
1.4\,GHz data from long Arecibo integrations used in \citet{ew01}. In our case, the
best fits slightly prefer small $\alpha <80^\circ$, although all values are allowed.
However, PSR J0659+1414 has weak but well measured emission extending well
beyond the peak of the pulse, giving a large $W_1 = 54^\circ$. Including this
constraint does indeed prefer near alignment. Our best fits are in fact for
$\alpha < 35^\circ$ although the large $\alpha > 130^\circ$ fits are nearly as good.
As the background to the constraints in this panel we show the gray scale
goodness of fit for the LAT data compared with a TPC model.
This model includes emission from below the null charge surface and so predicts
gamma-ray detections all the way to the spin axis.  There are in fact two broad zones
of reasonable TPC model fits consistent with the radio constraints near 
$\alpha \approx 40^\circ$ and $\alpha \approx 140^\circ$.  In contrast the OG model
for this pulsar's ${\dot E}$ has no $\gamma$-ray emission for solutions at 
$\alpha < 50^\circ$ or $\alpha > 130^\circ$. Thus, {\it if} we adopt both the
RVM and pulse width constraints, a large fraction of the remaining phase space is
incompatible with $\gamma$-ray emission from an OG model and a lower altitude
(eg. TPC-type) component is preferred. If we adopt the W1 constraint or the \citet{ew01}
$\alpha$ value this becomes a strong conclusion. However with the less strict
angle constraints inferred in this paper, OG exclusion is suggested, not required.

	The final parallax candidate is {\bf PSR J1932+1059} (B1929+10), which has
lower ${\dot E}$ than the cuts for our uniform sample. This pulsar
has been the subject of many polarization studies, summarized in \citet{ew01},
who find $\alpha=35.97\pm0.95^\circ$, $\zeta=61.52\pm1.3^\circ$ from a fit
excluding the main pulse, with the high S/N leading to very small statistical 
errors (blue circle).   For consistency with the other objects in this paper, 
we have fit the main peak width and position angle sweep.  Our RVM fit
prefers $\alpha < 60^\circ$, close to the EW01 value, while allowing all 
$\alpha$, as usual. However, given the large pulse width and low 
$h_{\rm LC}$ of Table 1, nearly aligned rotators are preferred. For 
the $W_{10}$ width we infer $\alpha<20^\circ$, while the $W_1$ width 
implies $\alpha<15^\circ$. The situation for this pulsar is not clear;
these pulse width constraints are not consistent with the EW01 fit. Moreover,
this pulsar has a widely separated faint pulse component which would
be identified as an interpulse for orthogonal solutions, but for 
more aligned solutions suggests a very wide pulse profile.  Such emission can 
only come from the open zone for very large $h_{LC}$, which we do not model here.  
Nevertheless, for either the EW01 solution or the RVM/pulse width constraints,
high altitude $\gamma$-ray emission is not expected to be visible at Earth for
this small ${\dot E}$ pulsar.  At first sight this would seem to provide a strong
confirmation of the connection between alignment and low $\gamma$-ray
flux.  However, with such a low ${\dot E}$, this pulsar is in the 
`death zone' where powerful $\gamma$-ray gap emission may have turned off.
\bigskip
\bigskip
\bigskip

\subsection{Sub-Luminous Candidates without Parallaxes}

	Since the objects with precise distances do not yet provide
a definitive test
of the nature of sub-luminous pulsars, we check other pulsars for which
the LAT provides relatively low luminosities at their estimated distances.
For example, {\bf PSR J0745$-$5353} (B0743$-$53) is assigned a distance of 0.25\,kpc in the
CL02 model, because it is superimposed on HII emission associated
with the Gum nebula. At this distance the flux is $>450\times$ less than that
expected from $L_{\gamma,heu}$. However, at the 7.1\,kpc distance implied by
the \citet{tc93} model the upper limit is not constraining (the pulsar does remain
sub-luminous for distances as large as 2\,kpc).
The combined RVM and pulse width constraints imply $\alpha > 150^\circ$ ($W_{10}$) or
$\alpha > 160^\circ$ ($W_{1}$).  Clearly, these
constraints indicate an anti-aligned rotator such that only $\gamma$-ray emission 
from below the null charge surface ($r<r_{\rm NC}$) should be visible.  Unfortunately the 
highly uncertain distance prevents us from drawing strong conclusions from 
the absence of flux from this pulsar. A parallax distance measurement
(or at least a lower limit) would be of particular value; if a low distance is 
confirmed it provides a sharp test of the $\gamma$-ray beaming geometry.

	Our final pulsar is {\bf PSR J1740+1000}. This pulsar is of
interest since it is young and is located well off the Galactic plane. This
makes for a very clean LAT flux limit and a relatively robust DM distance.
We have performed RVM fits using polarization data collected at Nan\c cay
with the BON pulsar back-end (see Theureau et al. 2011). 
The fits allow a large $\alpha$ range, but prefer
values near $120^\circ$. However, such an orthogonal rotator is very
difficult to accommodate with the very wide observed pulse, even for
the relatively large $h_{LC}=0.122$ inferred here from $\Delta \phi_{PA}$.
The standard pulse width constraints are shown in Figure 4;
to accommodate the $W_{10}$ width implies $\alpha < 70^\circ$ or $>120^\circ$, 
while the $W_1$ width requires $\alpha < 50^\circ$ or $>140^\circ$.
Interestingly, if we fit the numerical magnetosphere models, the large
$\alpha$ solution becomes preferable and in fact shifts to slightly higher values.
The red (dashed) contours show the numerical model fit including $\Psi$ points within
$W_{10}$.  For numerical fits placing points out to the $W_1$ pulse width 
in the open zone, the allowed region moves inside the magenta $W_1$ contour and
we find $\alpha \approx 30^\circ $ or $150^\circ$, where one would expect no high altitude
$\gamma$-ray emission at this ${\dot E}$. However, if we only consider the
data within the $W_{10}$ pulse width an appreciable region including
OG-type emission is allowed. Thus the geometry of this pulsar is not
yet sufficiently constrained to test the models. Improved orientation 
constraints, perhaps from additional radio and X-ray observations, 
are needed. A parallax measurement would also be very valuable for 
strengthening the use of this pulsar to test luminosity models.

\subsection{Other Nearby, Energetic Radio Pulsars}

	The DM-distance cut selects three additional young energetic radio pulsars,
but these, plotted on Figure 1b, may well have typical LAT pulsar luminosities. 
First, PSR J0834$-$4159 is undetected with a flux limit giving $\sim
1/3\,L_{\gamma,heu}$ at its DM distance.  
Like J0745$-$5353, this pulsar's DM distance estimate is dramatically 
smaller in the CL02 model (1.6\,kpc) than in the \citet{tc93} TC93 model (9.7\,kpc), due to
inclusion of nearby HII complexes.   However, unlike J0745$-$5353, even a 
modest factor of $2$ increase in the true distance would make the present
flux bound unconstraining. Geometrical data on this pulsar are limited.
It has pulse components separated by $\approx 165^\circ$, so it is a likely 
interpulsar \citep{wj08}. However, unlike other pulsars discussed in this
paper it has very little linear polarization, so the geometry has not been confirmed
by RVM modeling. In sum, this is not a strong case for a sub-luminous pulsar.
Indeed, the limited radio information suggests a near-orthogonal rotator and thus
$\gamma$ emission beamed toward Earth. Continued LAT exposure,
and improved radio data can help clarify the situation.

	Statistically significant LAT flux is found in the direction of PSR J0857$-$4424.
Like the other pulsars in this region, the DM-estimated distance had a major adjustment
and should be considered uncertain. In addition, background systematics can
perturb the LAT flux estimate. We conclude that this pulsar is likely
not sub-luminous, although a LAT pulsed detection is required for firm conclusions.
Unfortunately the lack of significant linear polarization in this pulsar
will make it very difficult to extract detailed radio geometry constraints.

	The last nearby energetic object is PSR J1722$-$3712 (B1719$-$37) at 1.9\,kpc (CL02).
This interpulsar has been the subject of a very careful polarization study by 
\citet{keith10} who find (EW01-corrected) angles $\alpha = 89.3\pm0.1^\circ$ 
and $\zeta= 83.9\pm0.3^\circ$. 
Thus we have high confidence that this is an orthogonal rotator and we
expect to see visible OG emission. The LAT does, in fact, provide an unpulsed detection.
There are no particular issues with DM in this direction and the LAT
detection is consistent with a point source, localized to the radio pulsar position
and having a pulsar-like $\gamma$-ray spectrum. The inferred luminosity is
quite consistent with $L_{\gamma,heu}$. Thus this object has a well constrained
geometry indicating that the outer magnetosphere $\gamma$-ray beams should be visible
and we do indeed detect the source. Interestingly, of the five $d<3$\,kpc interpulsars 
in the study of \citet{keith10} three are LAT detected (PSR J1057$-$5226=B1055$-$52 at 0.7\,kpc, 
PSR J1722$-$3712=B1719$-$37 at 1.9\,kpc and PSR J0908$-$4913=B0906$-$49 at 2.5\,kpc) while the other 
two (PSR J1549$-$4848 at 2.7\,kpc and PSR J1739$-$2903=B1736$-$29 at 2.5\,kpc)
have at present high upper limits or weak detections, quite consistent with 
$L_{\gamma,heu}$.

	LAT pulse searches for these objects and similar interpulsars with 
well constrained geometries can provide important model tests. Note
that with $\zeta \approx 90^\circ$ we are probing emission very near the null charge
surface for these objects. As it happens, the presence of $\gamma$-ray OG emission
in this region is sensitive to the magnetospheric currents \citep{hir06}. 
The shape and phase of the LAT pulsations allow us to trace the emission to
particular magnetosphere zones. When kinematic distances are also available, we
then have the actual luminosity of these zones, a particularly powerful
model constraint. Unfortunately such constraints are very difficult to obtain
for the nearly-aligned pulsars which are plausibly associated with the 
sub-luminous pulsar class discussed here.

\section{Conclusions and Future Prospects}

	We have examined the set of 12 young, energetic radio pulsars
with  $d\le 2$\,kpc. Most follow the standard LAT
pattern of powerful, efficient high altitude $\gamma$-ray emission
with luminosity $\sim L_{\gamma, heu}$. However PSR J0659+1414, at
1/20$^{\rm th}$ of this luminosity, is a clear outlier. Present flux
bounds make a good case that PSR J0538+2817 is also $\sim 10\times$
sub-luminous.  Although they lack parallaxes, PSRs J0745$-$5353 and J1740+1000
are also likely to produce $<0.1L_{\gamma, heu}$. PSR J0358+5413 has 
a parallax, but a less restrictive bound, at present. Thus we find
that 4 and perhaps 5 of our nearby energetic sample may be
members of this sub-luminous class, although for several of these pulsars
the case could be strengthened with parallax distance measurements.

	We have attempted to determine whether this may be attributed
to beaming away from Earth, as would be expected for high-altitude
(OG) emission and aligned or anti-aligned spin geometries. Unfortunately,
precise geometrical constraints are very difficult to obtain for
aligned rotators. We do find that for our sub-luminous pulsar candidates
the present radio constraints prefer aligned geometries. The best
cases are probably PSR J0538+2817 and PSR J0745$-$5353. Although it
lies below the ${\dot E}$ cut for our sample set, PSR J1932+1059
also seems sub-luminous and aligned. However, no one case for alignment
is compelling. 

	The converse argument is in somewhat better shape. When
we know that the pulsar is orthogonal, we seem to see $\gamma$-ray
emission at the expected $L_{\gamma, heu}$.  PSR J1722$-$3712 is
an excellent example. Although the undetected PSR J0834$-$4159 may 
be orthogonal, its distance estimate is particularly uncertain, and
its present flux limit is quite likely compatible with $L_{\gamma, heu}$.

	In the population synthesis of \citet{wr11} it was concluded 
that for OG geometries $\approx 30$\% of the radio-selected ${\rm Log}({\dot E}) > 33.5$ 
pulsars should be undetected in the $\gamma$-rays, simply due to beaming.
Our present sub-luminous pulsar fraction, 
(4 to 5)/12 pulsars, is consistent with this ratio, given the small number
statistics. However we would not expect many more sub-luminous pulsars
unless the radio beams are substantially larger than and/or
$\gamma$-ray beams are substantially smaller than assumed in these beaming
computations. The fact that we see evidence for high altitude ($h_{\rm LC}> 0.1$)
radio emission in several of our pulsars supports the presence of
wide radio beams and a somewhat larger sub-luminous pulsar fraction;
\citet{rmh10} have also argued that the statistics of radio and
$\gamma$-ray detections imply wide radio beams for young pulsars.

	The {\it detection} of PSR J0659+1414 at $\sim L_{\gamma, heu}/20$
presents an important addition to this beaming picture. This
low flux level, together with the $\gamma$-ray pulse's unusual
phase and soft spectrum suggest atypical magnetospheric emission. 
If this is an aligned (or anti-aligned) rotator, this must be low
altitude $r<r_{\rm NC}$ emission. A set of J0659+1414 type pulsars 
is certainly needed to probe the origin of this low flux.

	If we assume that several of our sub-luminous candidates join the
J0659+1414 class, we already have hints to the common features. Certainly
non-orthogonal geometries seem preferred, although better geometrical
constraints are needed for most sources to establish them as aligned rotators.
Other plausible peculiarities for PSR J0659+1414 exist; for example it 
has the lowest light cylinder field $B_{LC}$ of any LAT radio pulsar.	
Table 1 lists $B_{LC}$ for our candidates. No strong trend is evident,
and certainly the few kG fields of PSRs J0358+5413, J0538+2817
and J1740+1000 are quite typical of those of detected LAT pulsars.
For these objects, at least, orientation seems more promising.

	Since several of our sub-luminous candidates have limits on 
luminosity approaching the J0659+1414 level, it will
be important to see if increased LAT exposure or pulsed searches
can detect evidence of similar low level $\gamma$-ray emission. 
Any such pulse detections for these sources must then represent an
atypical `sub-luminous' mechanism.  The phasing of such pulses
can be used to cement the geometrical location and their spectrum and dependence
on spin properties should help lock down the emission physics. The prospect
of using the LAT to probe a second domain of pulsar particle acceleration, in
addition to the established high luminosity outer magnetosphere emittors, is very exciting.

\bigskip

	   The Fermi LAT Collaboration acknowledges generous ongoing support from 
a number of agencies and institutes that have supported both the development and 
the operation of the LAT as well as scientific data analysis. These include the 
National Aeronautics and Space Administration and the Department of Energy
in the United States, the Commissariat a l'Energie Atomique and the Centre 
National de la Recherche Scientifique/Institut National de Physique Nucleaire 
et de Physique des Particules in France, the Agenzia Spaziale Italiana and the 
Istituto Nazionale di Fisica Nucleare in Italy, the Ministry of Education,Culture,
Sports, Science and Technology (MEXT), High Energy Accelerator Research Organization 
(KEK) and Japan Aerospace Exploration Agency (JAXA) in Japan, and the K. A. Wallenberg
Foundation, the Swedish Research Council, and the Swedish National Space Board in Sweden.

   Additional support for science analysis during the operations phase is gratefully 
acknowledged from the Istituto Nazionale di Astrofisica in Italy and the Centre 
National d'Etudes Spatiales in France.

This paper has made extensive use of the ATNF pulsar catalog \citep{met05}.
This work was supported in part by NASA grants NNX10AD11G and NNX10AP65G.
Support for this work was also provided by the National Aeronautics and
Space Administration through Einstein Postdoctoral Fellowship Award Number
PF0-110073 issued by the Chandra X-ray Observatory Center, which is operated by
the Smithsonian Astrophysical Observatory for and on behalf of the National
Aeronautics Space Administration under contract NAS8-03060.

\end{document}